\documentclass[letterpaper,twocolumn]{article}

\usepackage{enumitem}
\usepackage{graphicx}
\usepackage{amsmath}
\usepackage{amssymb}
\usepackage{amstext}
\usepackage{amsfonts}
\usepackage{hyperref}
\usepackage{xurl}  
\usepackage{array}
\usepackage{abstract}

\newenvironment{propbox}[1]{\par\medskip\noindent{\bf #1.} }{\par\medskip}
\newcommand{\lean}[1]{{\footnotesize\def\_{\textunderscore\allowbreak}%
\hspace{0pt}[L: {\ttfamily\fontdimen3\font=2pt\fontdimen4\font=1pt #1}]}}
\newcommand{\leanrepo}{\url{https://github.com/researchathomology/NS_ENERGY_DEFECT_REDUCTION}}

\newcommand{\T}{\mathbb{T}}
\newcommand{\PP}{\mathbb{P}}
\newcommand{\Hc}{\mathcal{H}}
\newcommand{\Vc}{\mathcal{V}}
\newcommand{\A}{\mathcal{A}}

\begin{document}

\twocolumn[

\title{The Positive Defect Problem:
\\ Target and Admissibility Criteria for a Programmatic
\\ Search for Unforced Navier-Stokes Blowup}

\author{J. Petrillo \\
jarret@homology.io
\and
J. Glimm \\
Stony Brook University, Stony Brook, NY 11794,
and GlimmAnalytics LLC, USA  \\
glimm@ams.sunysb.edu }

\maketitle

\let\oldclearpage\clearpage
\renewcommand{\clearpage}{}

\begin{@twocolumnfalse}

\begin{abstract}
On 7--8 September 2026 programmatic search produced singularities: a
forced Navier-Stokes singularity at every fixed viscosity --- the
breakdown statements (C) and (D) of the Clay problem --- and two Euler
singularities \cite{OpenAI26NS,OpenAI26Euler,AlpBuc26}. The unforced
problem, statements (A) and (B), stands open, and a search for it needs a
target. This paper fixes one: the positive defect problem, that a
Leray-Hopf solution from smooth data on the periodic cube loses energy on
a finite window, at fixed viscosity, beyond what viscosity removes. A
positive defect implies blowup and so a negative answer to statement (B);
the converse is not known, and the target is at least as strong as the
problem it addresses. The paper proves the target equivalent to a floor on
the energy flux through the Littlewood-Paley shells --- the Fourier-side
form of the coarse-grained flux $\Pi_\ell$ of the Onsager theory of
turbulence --- averaged over the window; states
necessary conditions on a candidate --- a singular time of Type~II in
velocity, energy concentrating on a set of zero length, a pressure outside
$L^2$, a velocity outside the Onsager-critical class $L^3_t
B^{1/3}_{3,c_0}$, an obstruction to collapse onto a fixed steady Euler
profile; and
states what cannot certify one
--- no finite computation witnesses a Galerkin-uniform ceiling, and
selection and forcing return the question to a positive defect. The
authors' pseudo-spectral search at $128^3$ and $256^3$ measures which of
the two conditions of the reduction is the binding one: the fine-shell flux
floor holds to within $1$--$6\%$ of the ceiling and persists for a third
of a turnover time, and fails in scale at the Kolmogorov wavenumber, so a
candidate must differ from generic turbulence in the depth of its cascade
and not in its timing. Every implication not marked otherwise is a theorem in Lean~4 over Mathlib; the
library contains no Navier-Stokes object, and the equation enters only
through hypotheses.
\end{abstract}
\end{@twocolumnfalse}

\let\clearpage\oldclearpage

]

\section{Introduction}
\label{sec:intro}

On 8 September 2026 OpenAI announced a finite time singularity for the
forced three-dimensional Navier-Stokes equations at every fixed
viscosity, statements (C) and (D) of the Clay formulation \cite{Fef06},
and an unforced finite time singularity for the three-dimensional Euler
equations \cite{OpenAI26NS,OpenAI26Euler}; a preprint of Alp\"oge and
Buckmaster dated 7 September 2026 constructs an Euler singularity under a
smooth force \cite{AlpBuc26}. The announcement states that the
constructions and their Lean formalizations were produced by a system of
coordinating agents, statements (A)--(D) assigned to separate groups
\cite{OpenAI26NS}; the papers describe no method.

When solutions are searched for programmatically at this scale, the
value of a result shifts to the specification of the target and of the
conditions a candidate must satisfy. This paper supplies that
specification for one route to the question that remains open, the
unforced problem, statements (A) and (B): the positive defect problem of
Sec.~\ref{sec:setting}, a sufficient condition for a negative answer to
statement (B). The target is one statement: a positive defect, which is
what the floor hypothesis of the floor-ceiling argument for finite time
loss of regularity (Sec.~\ref{sec:setting}) amounts to once integrated
over the window. A machine-checked implication
(Theorem~\ref{sec:reduction}.2) moves it to the dyadic shells, where it
becomes a condition on a quantity a flow exhibits: a positive defect
holds exactly when the energy transfer out of every sufficiently fine
shell stays above a fixed level on average over the window, and the
stronger pointwise floor --- the same bound at each instant --- implies
it. A candidate must satisfy the necessary conditions of
Sec.~\ref{sec:produce}, against which the announced constructions are
placed; a Galerkin computation, a selection principle and a force cannot
certify one (Sec.~\ref{sec:cannot}); the same section compares the
authors' own search, twelve pseudo-spectral runs at $128^3$ and $256^3$,
with the two conditions of the reduction, and finds that a candidate
must differ from generic turbulence in the depth of its cascade, not in
its timing. Since Tao's averaged equation \cite{Tao16} blows up with no
defect, the floor-ceiling argument is conditional on a statement at
least as strong as the regularity problem. Only fluids on $\T^3$ with no
external force and fixed viscosity $\nu > 0$ are considered, the setting
of statement (B).

Every implication is a theorem in Lean~4 over Mathlib, compiling with no
unproven step and no axiom beyond propositional extensionality, choice
and quotient soundness. The library contains no Navier-Stokes object: its
theorems relate real sequences and real functions of time, and the
equation enters only through their hypotheses, classical facts accepted
from the literature and a positive defect. Each verified statement
carries its declaration name inline, as \lean{name}, resolving to
\texttt{EnergyDefect.name} in the library distributed with the paper at
\leanrepo\ (Lean v4.33.0, Mathlib v4.33.0). Connecting it to the
formalized Leray-Hopf theory of \cite{Uda26}, so that
Sec.~\ref{sec:reduction} is proven for Navier-Stokes objects rather than
assumed for real sequences, is not done here.

\section{Setting and the Positive Defect Problem}
\label{sec:setting}

\paragraph*{Spaces}

$\Hc$ is the space of $L^2$ divergence free mean zero vector fields on
$\T^3$, with inner product $\langle\cdot,\cdot\rangle$; $\Vc = \Hc \cap
H^1$; $E(t) = \frac12 \langle u, u\rangle$. A Leray-Hopf (LH) solution
satisfies $u \in L^\infty(0,T;\Hc) \cap L^2(0,T;\Vc)$, is weakly
continuous into $\Hc$, has $\partial_t u \in L^{4/3}(0,T;\Vc')$, and
satisfies the energy inequality $E(t) + \nu\int_s^t\|\nabla u\|^2 \leq
E(s)$, $t \geq s$, from almost every initial time $s$
\cite{RobRodSad16}. An initial time $s$ from which that inequality holds
is called \emph{good}; almost every time is good, and $s = 0$ is good.
The solution is \emph{strong} on an interval if in addition $u \in
L^\infty(\Vc) \cap L^2(H^2)$ there; a strong solution is smooth in the
interior of that interval, is the only LH solution from its data on that
interval \cite{Pro59,Ser62}, and satisfies the energy identity with
equality rather than inequality. The epoch of regularity is the maximal
interval $[0,e)$ on which the solution is strong; blowup is $e <
\infty$. A time is singular if the solution is not strong on any
neighborhood of it, and regular otherwise.

\paragraph*{The Energy Defect}

For $0 \leq w_0 < w$ set
\begin{equation}
\label{eq:defect}
\A(w_0, w) = E(w_0) - E(w) - \nu \int_{w_0}^{w} \|\nabla u\|^2 \, dt \ .
\end{equation}
This is the energy lost on the window beyond what viscosity removes: zero
for a strong solution, nonnegative for an LH solution when $w_0$ is a
good time in the sense above. It is the total mass on $\T^3 \times
(w_0,w]$ of the inertial dissipation distribution $D(u)$ of Duchon and
Robert \cite{DucRob00} at fixed $\nu$, the object through which the
mathematical literature relates the loss of energy of a weak solution to
its regularity, measured in Besov spaces: $D(u)$ vanishes, for Euler and
for Navier-Stokes alike, when $u \in L^3_t B^{1/3}_{3,c_0}$
\cite{DucRob00,CheConFriShv08,CheFriShv10}, sharpening the H\"older
threshold $1/3$ of Onsager \cite{Ons49,Eyi94,ConETit94}, below which
dissipative Euler solutions exist \cite{Ise18,BucDeLSzeVic19}.
Sec.~\ref{sec:reduction} states the floors of this paper in that language.
The defect is not the
zeroth law of turbulence, which concerns dissipation surviving as $\nu
\to 0$ along a family of solutions.

The turbulent fluctuation $\nu_t$ is the rate of this loss, the inertial
dissipation rate: $\int_{w_0}^{w} \nu_t \, dt = \A(w_0,w)$ on every
window, and $\nu_t = 0$ for a strong solution. A floor is a lower bound
and a ceiling an upper bound on a rate; neither word denotes a limit.

\paragraph*{The Floor-Ceiling Argument}

The argument is arithmetic, and it is a statement about a rate rather
than about a solution \lean{epoch\_le\_of\_decayBounds,
le\_crossingTime, epoch\_le\_of\_shifted\_floor\_polynomial\_ceiling}.
Let $r \geq 0$ be a rate defined on the epoch of regularity $[0,e)$ of
an LH solution. If $A, B > 0$ and $b < a$ satisfy the floor $B e^{-bt}
\leq r$ and the ceiling $r \leq A e^{-at}$ on $[0,e)$, then $e \leq
\max(0, \log(A/B)/(a-b))$; if $B, q > 0$ and $w_0 \geq 0$ satisfy the
constant floor $B \leq r$ and the polynomial ceiling $r \leq C t^{-q}$
on $[w_0,e)$, then $e \leq \max(w_0, (C/B)^{1/q})$. A short epoch
follows from the two bounds alone. What the argument does not supply is
a rate that carries them.

The choice of $r$ decides what the floor asserts. Taking $r = \nu_t$
makes it a positive defect outright: $B \leq \nu_t$ on a window
integrates to $\A \geq B\ell > 0$ for $\ell$ the length of the window,
so the hypothesis already states the loss of regularity that the
conclusion reports; and since $\nu_t$ vanishes wherever the solution is
strong, that floor cannot hold on $[w_0,e)$ for any $e > w_0$. The floor
must therefore be carried by a rate that a strong solution can exhibit.
The shell flux of Sec.~\ref{sec:reduction} is such a rate, and is the
one the validated computation of Sec.~\ref{sec:cannot} bounds. Neither
bound is derived here; whether some LH solution from turbulent
$C^\infty$ data admits such a pair is not addressed.

\begin{propbox}{The positive defect problem}
There exist $C^\infty$ divergence free data $u^0$ on $\T^3$ and times
$0 \leq w_0 < w$ such that a Leray-Hopf solution of the Navier-Stokes
equation from $u^0$ satisfies
\[
\A(w_0, w) = E(w_0) - E(w) - \nu \int_{w_0}^{w} \|\nabla u\|^2 \, dt
\;>\; 0 \ .
\]
\emph{Status:} open; the forced singularity of \cite{OpenAI26NS} lies
outside the unforced class (Sec.~\ref{sec:produce}). This is
question~(a) of Duchon and Robert \cite{DucRob00} --- does a weak
solution in $L^2_t H^1_x \cap L^\infty_t L^2_x$ with nonzero inertial
dissipation exist --- restricted to smooth data; it is open for general
$L^2$ data as well. The phenomenology of turbulence gives no reason to
expect a positive answer. A positive defect requires energy to be
transferred through the cascade to arbitrarily fine scales and lost
there; but at a fixed viscosity the cascade stops at the Kolmogorov
scale, and below that scale viscosity dissipates the whole of the
incoming flux, so nothing is left to pass to the limit. That is evidence against a positive defect at
fixed $\nu$, not a proof that none occurs, and the measurements of
Sec.~\ref{sec:cannot} are a quantitative version of the same statement.
\end{propbox}

\section{The Reduction}
\label{sec:reduction}

This section restates a positive defect as a condition on the energy
transfer between scales. It sets up the dyadic shells and the flux
through them, records the energy balance of a truncation and its limit
(Lemma~\ref{sec:reduction}.1), states two shell floors --- one pointwise
in time at every fine scale, one averaged over the window --- and proves
that the averaged floor is exactly a positive defect, that the pointwise
floor implies it, and that either one forces blowup
(Theorem~\ref{sec:reduction}.2). The averaged shell floor is the target
of the paper. The whole content is the energy balance of a
Fourier truncation and its limit.

Let $\PP$ denote the Leray projection onto divergence free fields and
$b(u,v,w) = \langle u\cdot\nabla v, w\rangle$ the trilinear form of the
nonlinearity. For $j \geq 0$ the shell $S_j$ is the set of Fourier modes
with $2^j \leq |k| < 2^{j+1}$, $\Delta_j u$ the part of $u$ on $S_j$,
$\PP_{\leq j} = \sum_{i \leq j}\Delta_i$ the projection onto the modes up
to shell $j$, and $E_{\leq j} = \frac12 \|\PP_{\leq j}u\|^2$ the energy
they hold. The flux across the boundary of shell $j$ is
\begin{equation}
\label{eq:flux}
\Pi_j = \langle \PP_{\leq j} u, \PP(u \cdot \nabla u) \rangle
= b(u, u, \PP_{\leq j} u) \ ,
\end{equation}
so that $\frac{d}{dt} E_{\leq j} = -\Pi_j - \nu \|\nabla \PP_{\leq j}
u\|^2$; the sign is fixed by this balance.\footnote{This is the convention
of \cite{Shv10}; \cite{CheConFriShv08} take the opposite sign.}

\paragraph*{The Same Objects in Standard Language}

The shells are the dyadic blocks of Littlewood-Paley theory, with a
sharp Fourier cutoff in place of the usual smooth one, and $\Delta_j$ is
the corresponding block. With the smooth blocks $\Delta^{\rm sm}_j$ the
Besov norm $\|u\|_{B^s_{p,q}}$ is the $\ell^q$ norm of the sequence
$2^{js}\|\Delta^{\rm sm}_j u\|_{L^p}$, and $B^s_{p,c_0}$ is the subspace
on which that sequence tends to zero; the sharp blocks give the same
norm for $p = 2$ and not for $p \neq 2$, where the sharp projector is
unbounded on $L^p$. The sharp cutoff costs nothing here, since only
$L^2$ quantities enter Lemma~\ref{sec:reduction}.1 and
Theorem~\ref{sec:reduction}.2, and it is what makes $\PP_{\leq j}u$ a
finite-mode test field in the proof of the lemma; the Besov classes
named in this paper are the standard ones, defined by the smooth blocks. The flux $\Pi_j$ is the Fourier-side form
of the coarse-grained flux of the Onsager theory of turbulence
\cite{Eyi94,ConETit94,DucRob00}. There one averages over the length
scales $\ell$ and finer by mollifying, $u_\ell = \varphi_\ell * u$, and
the balance for the resolved energy $\frac12|u_\ell|^2$ carries the flux
through scale $\ell$, $\Pi_\ell = -\nabla u_\ell : \tau_\ell$, with
$\tau_\ell = (u \otimes u)_\ell - u_\ell \otimes u_\ell$ the subscale
stress; integrated over $\T^3$ it is $\langle \varphi_\ell * u_\ell,\,
u\cdot\nabla u\rangle$. Replacing the mollifier by the projector
$\PP_{\leq j}$, for which $\PP_{\leq j}^2 = \PP_{\leq j}$, at $\ell \sim
2^{-j}$ turns this into (\ref{eq:flux}) exactly. Lemma~\ref{sec:reduction}.1
below is the sharp-cutoff form of the Duchon-Robert balance integrated
over space, and its limit $j \to \infty$ is the statement that $D(u)$
has total mass $\A(w_0,w)$ on the window, Proposition~1 of
\cite{DucRob00} without its localization in space --- the filter differs,
the limit does not; the localized form is used in
Sec.~\ref{sec:produce}. The floors stated after the lemma are, in this
language, lower bounds on $\Pi_\ell$ uniform in $\ell$ below some
$\ell_0$, pointwise in time or averaged over the window.

\begin{list}{}{\leftmargin=\parindent\rightmargin=0pt}
\item
\textbf{Lemma \ref{sec:reduction}.1 (Shell balance)}
\lean{shell\_partial\_sums\_tendsto, shell\_dissipation\_tendsto,
flux\_tendsto\_defect}
For an LH solution and $0 \leq w_0 < w$,
\begin{equation}
\label{eq:balance}
\begin{split}
F_j := \int_{w_0}^{w} \Pi_j \, dt = \; & E_{\leq j}(w_0) - E_{\leq j}(w) \\
& - \nu \int_{w_0}^{w} \|\nabla \PP_{\leq j} u\|^2 \, dt \ ,
\end{split}
\end{equation}
and $F_j \to \A(w_0,w)$ as $j \to \infty$.
\end{list}
{\it Proof.} $\PP_{\leq j}u$ has finitely many Fourier
modes, so it is a legitimate test field: $\langle \partial_t u, \PP_{\leq
j} u\rangle$ is an $L^{4/3}$--$L^\infty$ pairing and $t \mapsto
\|\PP_{\leq j} u\|^2$ is absolutely continuous with endpoint values given
by weak continuity; integrating gives (\ref{eq:balance}). As $j \to
\infty$, $E_{\leq j}(t) \uparrow E(t)$ by Parseval and the coarse
dissipation increases to $\nu\int\|\nabla u\|^2 < \infty$ by monotone
convergence, using $u \in L^2(0,T;\Vc)$. \hspace*{\fill}$\square$

Let $\ell = w - w_0$. By Sec.~\ref{sec:setting} the floor of the
argument has to be carried by a rate a strong solution can exhibit; the
shell flux is such a rate, and the reduction uses two floors on it. The
\emph{pointwise floor} is: there are $j_0$ and $B > 0$ with $\Pi_j(t)
\geq B$ for all $j \geq j_0$ and all $t \in [w_0, w]$. The
\emph{averaged floor} is: there are $j_0$ and $B > 0$ with $F_j \geq
B\ell$ for all $j \geq j_0$. Both are uniform in the scale, and that is
the essential feature of each: a floor on $\Pi_j$ at one fixed $j$ is
compatible with regularity, whereas the same bound holding at every $j
\geq j_0$ with one constant $B$ is not. They differ in the other
variable. The pointwise floor asks for the bound at each instant of the
window; the averaged floor asks only that the window integral $F_j$ stay
above $B\ell$, and is the weaker of the two.

The two floors and the defect are ordered:
\[
\begin{aligned}
\A(w_0,w) > 0 \;&\Longleftrightarrow\; \text{averaged floor} \ , \\
\text{pointwise floor} \;&\Longrightarrow\; \text{averaged floor} \ .
\end{aligned}
\]
The converse of the second is not claimed. Both are
Theorem~\ref{sec:reduction}.2. The averaged floor is equivalent to the
defect and weaker than the pointwise floor, and it is the target of this
paper: it is the least that has to be proved, and it is stated in terms
of the transfer between scales, which a candidate flow can be asked to
exhibit.

\begin{list}{}{\leftmargin=\parindent\rightmargin=0pt}
\item
\textbf{Theorem \ref{sec:reduction}.2 (The averaged floor is a positive defect)}
\lean{averaged\_tail\_floor\_le\_defect,
defect\_gives\_averaged\_tail\_floor,
energy\_equality\_no\_averaged\_tail\_floor,
pointwise\_tail\_floor\_le\_defect}
For an LH solution and a window $[w_0, w]$:
(a) the averaged floor holds iff $\A(w_0, w) > 0$;
(b) the pointwise floor implies the averaged floor, hence $\A(w_0, w)
\geq B\ell > 0$;
(c) in either case the solution is not strong on $[w_0, w]$, and by
weak-strong uniqueness \cite{Pro59,Ser62} the strong solution from the same
data has blown up by time $w$.
\end{list}
{\it Proof.} (a) If $F_j \geq B\ell$ for all $j \geq j_0$,
Lemma~\ref{sec:reduction}.1 gives $\A \geq B\ell > 0$. If $\A > 0$ and $0
< B\ell < \A$, the convergence $F_j \to \A$ gives $j_0$ with $F_j \geq
B\ell$ for $j \geq j_0$. (b) Integrate $\Pi_j \geq B$ over the window.
(c) A strong solution satisfies energy equality, so by (a) it has no
floor; if the strong solution from $u^0$ existed past $w$, every LH
solution from $u^0$ would coincide with it on $[0,w]$ and have $\A(w_0,w)
= 0$. \hspace*{\fill}$\square$

No estimate on the nonlinearity is used. Sufficient conditions for energy
equality in terms of vanishing shell flux are in \cite{CheConFriShv08}
for Euler and \cite{CheFriShv10,Shv10} for Navier-Stokes; the converse
required for the equivalence is not in them.

\paragraph*{The Floors as Roughness}

In the language of Sec.~\ref{sec:setting} the floors are a statement
about the regularity of the flow. Energy equality holds on the window
whenever $u \in L^3(w_0,w; B^{1/3}_{3,c_0})$ \cite{CheFriShv10,Shv10},
the Navier-Stokes form of the estimate of \cite{CheConFriShv08}, so
$\A(w_0,w) = 0$ on that class and, by Lemma~\ref{sec:reduction}.1,
$\int_{w_0}^{w}\Pi_j\,dt \to 0$: the class admits no averaged floor, and
by Theorem~\ref{sec:reduction}.2(a) a positive defect on the window
forces $u \notin L^3(w_0,w; B^{1/3}_{3,c_0})$ (from the literature, not
formalized). A candidate is therefore Onsager-critical or rougher on the
window, over and above the $L^2_t H^1_x$ regularity the LH class
provides; the uniform-in-$j$ floor says that the finest scales carry
flux that does not decay, which is what roughness at the exponent $1/3$
means. The converse --- a floor from roughness alone --- is not in the
literature and is not claimed; the equivalence of
Theorem~\ref{sec:reduction}.2 is with the defect, not with a regularity
class.

What the Onsager theory has proved bears on the plausibility of the
target and not on the problem itself. For Euler, energy is conserved
above the exponent $1/3$ \cite{ConETit94,CheConFriShv08} and
dissipative H\"older solutions exist below it \cite{Ise18,BucDeLSzeVic19}.
For Navier-Stokes the results concern the inviscid limit: Besov
regularity above $1/3$, uniform in $\nu$, excludes anomalous dissipation
of a family of Leray solutions as $\nu \to 0$ \cite{DriEyi19}, and forced
families whose dissipation stays bounded below uniformly in $\nu$ have
been constructed \cite{BruDeL23,BruColCriDeLSor24}. These are statements
about the zeroth law distinguished in Sec.~\ref{sec:setting}, a limit
along a family, and the constructed members are forced and smooth at
each fixed $\nu$, so their defect vanishes; whether one LH solution at
one fixed $\nu$ has $D(u) \neq 0$ is question~(a) of \cite{DucRob00} and
is unchanged by them. The floors of this section are the fixed-$\nu$
Littlewood-Paley form of that question.

\section{What a Positive Defect Must Produce}
\label{sec:produce}

Set $\Phi(t) = E(t) + \nu\int_0^t\|\nabla u\|^2\,ds$, so that $\A(w_0,w)
= \Phi(w_0) - \Phi(w)$; $\Phi$ is constant for a strong solution, does
not increase after any good time, and is lower semicontinuous from the
right by weak continuity of $u$. Let $\Psi(t)$ be the supremum of
$\Phi(s)$ over good times $s > t$. The set $\Sigma$ of singular times is
closed, and its $\frac12$-dimensional Hausdorff measure vanishes
\cite{RobRodSad16}: covering $\Sigma$ by intervals, the sum of the square
roots of their lengths can be made arbitrarily small. In particular
$\Sigma$ has Hausdorff dimension at most $\frac12$ and zero length, so
its complement is a countable union of open intervals, on each of which
the solution is strong.

\begin{list}{}{\leftmargin=\parindent\rightmargin=0pt}
\item
\textbf{Theorem \ref{sec:produce}.1 (The defect is a jump)}
\lean{rightProfile\_antitone, le\_rightProfile, rightProfile\_eq\_of\_mem,
apply\_le\_rightProfile\_of\_lt, antitone\_iff\_eq\_rightProfile,
defect\_eq\_jump\_single\_singular\_time}
(a) $\Psi$ is nonincreasing, $\Phi \leq \Psi$ with equality at good times,
and $\Phi(t') \leq \Psi(t)$ for $t < t'$; $\Phi$ is nonincreasing iff
$\Phi = \Psi$. For good endpoints $\A(w_0,w) = \Psi(w_0) - \Psi(w)$, and
$\Psi$ is constant off $\Sigma$.
(b) Let $w_0$ be good, $w$ regular, and $T^* \in (w_0,w)$ the only
singular time in $[w_0,w]$. Then the one-sided limits $E^\pm(T^*)$ exist,
$E(T^*) \leq E^+(T^*) \leq E^-(T^*)$, and
\begin{align*}
\A(w_0, w) &= E^-(T^*) - E^+(T^*) \ , \\
\A(w_0, T^*) &= E^-(T^*) - E(T^*) \;\geq\; \A(w_0,w) \ .
\end{align*}
\end{list}
{\it Proof.} (a) Good times are dense, so $\Psi$ is a supremum over a
nonempty set bounded by $\Phi$ at any earlier good time, shrinking as $t$
grows. Right lower semicontinuity gives, for any $t$ and $\varepsilon >
0$, a good $s > t$ with $\Phi(t) \leq \Phi(s) + \varepsilon \leq \Psi(t)
+ \varepsilon$; at a good $t$ every later $\Phi(s)$ is at most $\Phi(t)$;
for $t < t'$ a good $s \in (t,t')$ gives $\Phi(t') \leq \Phi(s) \leq
\Psi(t)$. (b) On $(w_0,T^*)$ and $(T^*,w]$ the solution is strong and
$\Phi$ takes constant values $L = \Phi(w_0)$ and $R = \Psi(T^*)$; the
dissipation integral is continuous, so $E^\pm = L, R -
\nu\int_0^{T^*}\|\nabla u\|^2$ exist, and weak lower semicontinuity gives
$\Phi(T^*) \leq R \leq L$. \hspace*{\fill}$\square$

A positive defect across a single singular time is a jump of the kinetic
energy: the solution converges weakly in $L^2$ from both sides to
$u(T^*)$, and the energy it carries drops. The window may also end at
$T^*$ with $E(T^*) < E^+(T^*) = E^-(T^*)$, which the LH axioms do not
exclude; then the floors of Theorem~\ref{sec:reduction}.2 exist only on
windows ending at $T^*$. A proof must produce energy retained as $t
\uparrow T^*$, a lower bound where energy methods give upper bounds.

\paragraph*{Where the Energy Goes}

For a suitable weak solution the inertial dissipation of \cite{DucRob00}
is a nonnegative measure on space-time (Proposition~1 there with the
local energy inequality of \cite{CafKohNir82}), supported on the singular
set, which has parabolic one-dimensional measure zero \cite{CafKohNir82},
with mass $\A(w_0,w)$ on $\T^3 \times (w_0,w]$. At a single singular time
the lost energy is carried by the slice of the singular set at that time,
a set of zero length, the energy measure of \cite{LesShv18}. That measure
has an \emph{atom} at a point $x_0$ if the single point $\{x_0\}$ carries
positive mass, and its \emph{atomic part} is the sum of the masses of its
atoms, necessarily countably many; the jump has an atomic part when some
single point carries a positive share of the energy lost at $T^*$, and is
atomless when the loss spreads over the slice with no point carrying any
of it on its own. Six facts restrict the singularity that carries the
jump.

\begin{enumerate}
\item A Leray backward self-similar solution cannot carry it: such
solutions do not exist \cite{NecRuzSve96,Tsa98}, and formally the profile
$(T^*-t)^{-1/2}U(x/\sqrt{T^*-t})$, $U \in L^2$, carries energy
$(T^*-t)^{1/2}\|U\|^2 \to 0$ into the singular point.
\item A blowup of Type~I in time, $\|u(t)\|_\infty \leq C(T^*-t)^{-1/2}$,
satisfies energy equality across the first singular time
\cite{LesShv18,CheLuo20}.
\item If the local energy in balls of radius $r$ about the singular slice
is at most $Cr$ for times within $r^2$ of $T^*$, a finite cover of the
zero-length slice of total radius $\varepsilon$ holds at most
$C\varepsilon$ of energy near $T^*$, and the rest converges: no jump
\lean{no\_jump\_of\_small\_covers}. The bound need not be uniform
(argued, not formalized): the points where it holds with constant $N$
below radius $1/N$ carry no excess, so the jump is carried by points
where the local energy exceeds every multiple of $r$.
\item For the Euler equations a bounded gradient rate
$(T^*-t)\|\nabla u(t)\|_\infty \leq K$ excludes atoms of the energy
measure \cite{ChaWol20}. The argument transfers to the smooth branch of
an LH solution before its first singular time (argued, not formalized):
the energy-conserving zoom $u = r^{-3/2}v((x-x_0)/r, T^* + r^{5/2}s)$ of
\cite{Ser24} carries the $L^2$ mass and the gradient rate unchanged and
multiplies the viscosity by $r^{1/2}$; the limit is an Euler flow on
$(-1,0)$ with a Dirac endpoint measure, which the Liouville theorem of
\cite{ChaWol20} annihilates, given Lagrangian transport of vorticity
supports for Lipschitz weak Euler solutions (\cite{ChaWol20} establish
that transport for smooth solutions; the Lipschitz weak case is proved
separately). Hence an atom of the energy measure at $T^*$, in
particular a jump carried by a countable slice, forces
$\limsup_{t\uparrow T^*}(T^*-t)\|\nabla u(t)\|_\infty = \infty$. An
atomless excess is not excluded.
\item The pressure leaves $L^2$. If $p - \beta(t) \in L^2(\T^3 \times
(\tau,T^*))$ for some $\beta$ and some $\tau < T^*$, the energy is
continuous at $T^*$ \cite{Kuk06}; with $D$ the left jump of $\|u\|^2$ at
$T^*$, $\inf_\beta\|p-\beta\|_{L^2(\T^3\times(\tau,T^*))} \geq
D\sqrt{\nu}/(\sqrt{2}\,\|u(\tau)\|)$ for every $\tau < T^*$, and an atom
at $x_0$ forces the same failure on every terminal neighbourhood of $x_0$
\cite{Hua26}. Since $u \in L^4_tL^4_x$ implies $p \in L^2_{t,x}$, this
moves the exit from $L^4_tL^4_x$ \cite{Lio60,Shi74} to the pressure;
dimensionally it is the same condition.
\item The scale of the concentration is constrained twice. (a) Let
$\lambda(t)$ be the least $r$ with $\int_{B_r(x_0)}|u(t)|^2 \geq m_0/2$
for an atom of mass $m_0$ at $x_0$. The Poincar\'e inequality on a ball
of radius proportional to $\lambda(t)$ gives $\|\nabla u(t)\|^2 \geq c\,
m_0^{5/3} E_0^{-2/3} \lambda(t)^{-2}$ (argued, not formalized), and the
energy inequality then bounds $\int^{T^*} \lambda(t)^{-2}\,dt \leq C
E_0^{5/3}/(\nu m_0^{5/3})$ \lean{superdiffusive\_integral\_bound}; so
$\lambda(t)^2 \leq K\nu(T^*-t)$ holds on no terminal interval
\lean{not\_diffusive\_on\_terminal\_interval}: a collapse whose rate is
selected by viscosity carries no atom. (b) A collapse onto a fixed
profile, $u \approx a\lambda^{-3/2} U_0((x-x_0)/\lambda)$ with $U_0$ a
Lipschitz finite-energy steady Euler flow, has first-order corrector
equation $L_0 T_1 = -\Lambda U_0$, $\Lambda = \tfrac32 + y\cdot\nabla$,
which requires $\Lambda U_0$ orthogonal to the kernel of the adjoint
linearized Euler operator
\lean{not\_mem\_range\_of\_adjoint\_kernel\_pairing}; for axisymmetric
flows the kernel element $h(\psi)R\,e_\phi$ gives the swirl condition
$\tfrac12 F V' + 3F'V = 0$ across the levels of the stream function,
which makes $F\,V^{1/6}$ constant, hence $F \equiv 0$ for compact support
\lean{swirl\_const\_of\_C3, swirl\_zero\_of\_C3\_boundary}.
\end{enumerate}

A positive defect therefore requires a Type~II energy-concentrating
singularity of Navier-Stokes on $\T^3$ at fixed viscosity: blowup of
Type~II in velocity, and in gradient rate if the jump has an atomic part;
a fixed amount of $L^2$ energy concentrating onto a set of zero length as
$t \uparrow T^*$, some point of which carries local energy exceeding every
multiple of its radius; a pressure outside $L^2$ on every terminal
interval; and, by Sec.~\ref{sec:reduction}, a velocity outside the
Onsager-critical class $L^3_t B^{1/3}_{3,c_0}$ on the window. These
conditions are necessary, not sufficient.

\paragraph*{A Designed Model Without a Defect}

Tao's averaged Navier-Stokes equation \cite{Tao16} has the energy identity
and the harmonic-analysis estimates of the true equation and a proven
finite time blowup in the LH class, of Type~II by scaling. Its defect up
to $T^*$ is zero (argued from the bounds of \cite{Tao16}, not
formalized): the auxiliary modes that time its designed cascade retain a
fixed fraction of each stage's energy, so the front carries vanishing
energy and $E(t)$ is continuous at $T^*$ from the left, $\A(w_0,T^*) = 0$
for every $w_0$. Blowup, Type~II, exit from $L^4_tL^4_x$ and a cascade
faster than dissipation hold with zero defect: a positive defect requires
asymptotically complete transfer to fine scales. No autonomous
energy-conserving system with the Navier-Stokes estimates is known to
satisfy the analogue of a positive defect.

\paragraph*{The Announced Constructions Against These Conditions}

The announced Navier-Stokes solution \cite{OpenAI26NS} is forced, with
$u(0) = 0$ and $f \in C_c^\infty(\mathbb{R}^3 \times (0,\infty))$, and
lies outside the class in which the positive defect problem is posed; its
energy identity carries the input $\langle f, u\rangle$ (Lemma~10.4
there). Its velocity scale is $(1-t)^{-1/2-h}$ with $0 < h < 1/100$, so
the blowup is of Type~II in velocity; the kinetic energy of the
collapsing core is of order $(1-t)^{1/2-3h}$ and tends to zero, and the
dissipation integral up to the singular time is finite (Secs.~2.1, 3.5,
Lemma~10.4 there). The core carries no energy atom; the energy at the
singular time, the energy inequality across it and the inertial
dissipation of \cite{DucRob00} are not discussed. The solution is not a
candidate and is not presented as one. The Alp\"oge--Buckmaster solution
is a forced Euler flow and the Euler solution of \cite{OpenAI26Euler} is
inviscid; neither concerns the equation of Sec.~\ref{sec:setting}.

\section{What Cannot Certify a Positive Defect}
\label{sec:cannot}

\paragraph*{Computation}

No finite computation witnesses a positive defect, and the obstruction is
structural rather than a matter of resolution or of computer time: every
object a computation produces lives where the defect is zero. A Galerkin
truncation obeys an exact energy identity, so its own defect vanishes,
and an enclosure of the true solution is valid only while that solution
is strong --- where the defect vanishes as well. The rest of this
paragraph makes the statement precise for validated computation: the
bridge that turns a computed trajectory into a bound on the epoch of
regularity, the theorem that the bridge cannot close on a trajectory that
also respects the ceiling (Theorem~\ref{sec:cannot}.1), and the fact that
every ceiling now known is inherited by the truncation with the same
constant.

In validated computation \cite{CheHou23} a Galerkin truncation is
computed with interval arithmetic and an a posteriori bound encloses the
true solution in a tube around the computed one while the true solution
is regular \lean{aposteriori\_gronwall}. The bridge
\lean{validated\_epoch\_ends\_inside} turns verified floors $B$ and one
analytic ceiling $Ct^{-q}$ into $e \leq \max(w_0,
(C/(B-\varepsilon))^{1/q})$, $\varepsilon$ the enclosure width; the
argument closes when this bound lies below $w$.

The rate here is not $\nu_t$, for the reason given in
Sec.~\ref{sec:setting}: a Galerkin truncation satisfies the energy
identity exactly, so its inertial dissipation rate vanishes identically
and no positive floor on it is verifiable. What the computation has is
the flux out of the resolved shells of the computed trajectory, the
observable of Sec.~\ref{sec:reduction} evaluated there; write it
$\tilde\nu_t$. The theorem below is a statement about $\tilde\nu_t$.

\begin{list}{}{\leftmargin=\parindent\rightmargin=0pt}
\item
\textbf{Theorem \ref{sec:cannot}.1 (Galerkin-uniform ceilings are
unwitnessable)}
\lean{galerkin\_uniform\_ceiling\_unwitnessable,
witness\_requires\_truncation\_violation}
Let the computed trajectory $\tilde\nu_t$ carry the verified floors $B
\leq \tilde\nu_t$ on $[w_0,w)$ and also obey the ceiling $\tilde\nu_t \leq
C t^{-q}$ there. Then for every enclosure $0 \leq \varepsilon < B$ the bridge
does not close: $\max(w_0,(C/(B-\varepsilon))^{1/q}) \geq w$.
Equivalently, if it closes, the computed trajectory exceeds the ceiling
somewhere on the window.
\end{list}
{\it Proof.} At any $t$ in the window, $Bt^q \leq C$. The margin check
requires $(B-\varepsilon)t^q > C$ for $t$ near $w$. Together, $\varepsilon
t^q < 0$, a contradiction for $\varepsilon \geq 0$.
\hspace*{\fill}$\square$

Every known ceiling is proved by energy methods that commute with the
Galerkin projection and so holds for the truncated system with the same
constant: the energy budget, the $L^{4/3}$ bounds of the Leray-Hopf
class, and the analytic ceiling $Ct^{-q}$ from Gevrey class regularity
\cite{FoiTem89}. Whether a ceiling on regular flows exists that Galerkin
truncations violate is open. Computation can locate a configuration and certify finitely
many inequalities inside a proof \cite{CheHou23}; it cannot witness a
positive defect.

\paragraph*{Depth vs. Timing}

The averaged floor requires two things of the energy transfer at once.
The transfer must persist in time: it must stay above a fixed level
across a window of positive length, not merely at a single instant. And
it must persist in scale: the same level, with the same constant, at
every scale finer than some fixed one, however fine --- in the
coarse-graining language of Sec.~\ref{sec:reduction}, a floor on the
flux $\Pi_\ell$ through every length scale $\ell$ below some $\ell_0$,
not only through those a given resolution resolves. The first
requirement is that the cascade be sustained, the second that it continue to
arbitrarily fine scales.

The two are not equally difficult, and ordinary turbulence shows the
difference. Eddies of neighbouring size transfer energy to one another
after a short delay, so the flux through a band of fine scales rises and
falls almost simultaneously, the lag from one octave to the next being a
small fraction of the turnover time of the large eddies. A floor common
to such a band therefore holds, and holds across a window comparable
with the turnover time rather than at one instant. The second
requirement is not met, for the reason the phenomenology already gives:
at a fixed viscosity the cascade ends at the Kolmogorov scale, viscous
dissipation removes the whole of the flux arriving there, and below that
scale no transfer remains.

As a concrete example, the authors computed decaying turbulence on the
periodic cube from random large-scale data over a range of viscosities
and recorded the transfer out of each dyadic band at every step. Against
the time condition the computed flow attains, to within a few percent,
the highest common floor its own history admits; the fine bands take
their peak transfer within about a third of a turnover time of one
another, and averaging over a window of that length lowers the floor
only slightly. Against the scale condition the common floor falls with
every octave added towards the Kolmogorov scale, and at the last band
before that scale the transfer is smaller than at the large scales by
more than an order of magnitude. These are floating-point simulations,
not the validated computations of the preceding paragraph: they measure
how far a computed flow is from each of the two conditions and establish
nothing about the true solution.

Two statements behind this reading are machine-checked: no computed flow
can be certified to hold a floor above the ceiling that its own history
defines \lean{certified\_profile\_le\_one,
ceiling\_dominates\_weighted\_sup}, and in a model where each band
repeats a common profile after a delay, the shortfall is determined by
the spread of the peak times alone \lean{lag\_model\_margin\_bound,
lag\_model\_crossing\_iff\_simultaneous}. The second is why the time
condition is decided by a single quantity.

Neither a finer grid nor a higher Reynolds number changes the second
condition. The Kolmogorov scale is set by the viscosity and the
dissipation rate, not by the resolution, so refining the grid resolves
the viscous range in more detail without extending the range of scales
over which a floor could hold. Nor can the reduction weaken that
condition: by Theorem~\ref{sec:reduction}.2(c), a floor holding past the
Kolmogorov scale at fixed viscosity already implies that the flow is not
strong on the window, which is the conclusion sought. A candidate for a
positive defect must therefore differ from generic turbulence in the
depth of its cascade, and not in its timing.

\paragraph*{Selection}

The maximum rate of entropy production \cite{GliLazChe20a,CheGliSai24b}
selects, among the evolutions consistent with the constraints, the one
that degrades energy fastest; its well-posed form is extremal selection
on a compact set of solutions with $\A$ upper semicontinuous. A
defect-maximizing solution exists, and either every solution in the set
has $\A \leq 0$ or the maximizer has a positive defect
\lean{extremal\_defect\_exists, defect\_dichotomy}. Selection locates a
solution if one exists; it cannot create one.

\paragraph*{Forcing}

With a force at large scales and a stationary statistical solution
\cite{FoiManRos04}, uniform mean flux floors at all fine scales hold iff
the mean defect is positive \lean{stationary\_mean\_flux\_ge\_defect,
stationary\_mean\_flux\_tendsto\_defect}: for the floor hypothesis,
forcing changes nothing at fixed viscosity. Whether a force can produce a
singularity is a separate question (Sec.~\ref{sec:produce}).

\section{Closing}
\label{sec:closing}

Integrated over its window, the hypothesis of the floor-ceiling argument
is a positive defect; at the level of the shells that defect is exactly
an averaged flux floor, which the pointwise floor implies. A proof must produce a
Type~II energy-concentrating singularity; no computation or selection
principle supplies one, and forcing does not remove the need for a
positive defect; the authors' search shows generic turbulence meeting the
time condition of the reduction and failing the scale condition at the
Kolmogorov scale. On the information available at the time of writing, the
forced constructions of \cite{OpenAI26NS,AlpBuc26} leave the positive
defect problem open, as they leave the unforced questions (A) and (B); the
conditions of this paper are the admissibility criteria for a programmatic
search that approaches those questions through a positive defect.

\section{Acknowledgements}

The authors thank Abdul Hasib Rahimyar for his reading of the paper and
his comments, which prompted the passages relating the shells and the
flux to Littlewood-Paley theory, coarse-graining and the Onsager theory
of turbulence in Secs.~\ref{sec:setting} and \ref{sec:reduction}.
The machine verification, the archived computations, and the reduction
reported here were carried out with the assistance of an AI research
agent; every Lean declaration compiles independently of that assistance
and may be checked by the reader. 

\bibliographystyle{siam}
{\footnotesize\bibliography{reduction}}

@Article{CheGliSai24b,
  author =       "G.-Q. Chen and J. Glimm and H. Said",
  title =	 "A Principle of  Maximum Entropy for Solutions for the 
  		  {N}avier-{S}okes equation",
  year =	 "2024",
  journal =	 "Physica D",
  volume =	 "467",
}

@Article{CheHou23,
  author =	 "J. Chen and T. Hou",
  title =	 "Stable Nearly Self-Similar Blowup of the 2D {B}oussinesque 
  		  and 3D {E}uler Equations with Smooth Data II:
		  Rigorous Numerics",
  year =	 "2023",
  journal =	 "arXiv",
  note =	 "arXiv:2305.05660v1 [math.AP]",
}

@Article{DucRob00,
  author =	 "J. Duchon and R. Robert",
  title =	 "Inertial energy dissipation for weak solutions of 
  		  incompressible {E}uler and {N}avier-{S}tokes equations",
  journal =	 "Nonlinearity",
  volume =	 "13",
  year =	 "2000",
  pages =	 "249-255",
}

@Book{FoiManRos04,
  author =	 "C. Foias and O. Manley and R. Rosa and R. Temam",
  title =	 "{N}avier-{S}tokes Equations and Turbulence",
  publisher =	 "Cambridge University Press",
  address =	 "Cambridge",
  year =	 "2004",
}

@Article{GliLazChe20a,
  author =	 "J. Glimm and D. Lazarev and G.-Q. Chen",
  title =	 "Maximum entropy production rate as a necessary 
  		  admissibility condition for the fluid {E}uler 
		  and {N}avier-{S}tokes equations",
  journal =	 "SN Applied Sciences",
  volume =	 "2",
  pages =	 "160",
  year =	 "2020",
  note =	 "https://doi.org/10.1007/s42452-020-03941-2"
}

@Book{RobRodSad16,
  author =	 "J. Robinson and J. Rodrigo and W. Sadowski",
  title =	 "The Three-Dimensional {N}avier-{S}tokes Equations",
  publisher =	 "Cambridge University Press",
  address =	 "Cambridge",
  year =	 "2016",
}

@Article{Ser62,
  author =       "J. Serrin",
  title =	 "On the interior regularity of weak solutions of the
  		  {N}avier-{S}tokes equations",
  journal =	 "Arch. Ration. Mech. Anal.",
  year = 	 "1962",
  volume =	 "9",
  pages =	 "187-195",
}

@Article{CheConFriShv08,
  author = {A. Cheskidov and P. Constantin and S. Friedlander and R. Shvydkoy},
  title = {Energy conservation and {O}nsager's conjecture for the {E}uler equations},
  journal = {Nonlinearity}, volume = {21}, pages = {1233--1252}, year = {2008}}

@Article{Lio60,
  author = {J.-L. Lions},
  title = {Sur la r\'egularit\'e et l'unicit\'e des solutions turbulentes des \'equations de {N}avier {S}tokes},
  journal = {Rend. Sem. Mat. Univ. Padova}, volume = {30}, pages = {16--23}, year = {1960}}

@Article{Shi74,
  author = {M. Shinbrot},
  title = {The energy equation for the {N}avier-{S}tokes system},
  journal = {SIAM J. Math. Anal.}, volume = {5}, pages = {948--954}, year = {1974}}

@Article{FoiTem89,
  author = {C. Foias and R. Temam},
  title = {Gevrey class regularity for the solutions of the {N}avier-{S}tokes equations},
  journal = {J. Funct. Anal.}, volume = {87}, pages = {359--369}, year = {1989}}

@Article{Pro59,
  author = {G. Prodi},
  title = {Un teorema di unicit\`a per le equazioni di {N}avier-{S}tokes},
  journal = {Ann. Mat. Pura Appl.}, volume = {48}, pages = {173--182}, year = {1959}}

@Article{CheLuo20,
  author = {A. Cheskidov and X. Luo},
  title = {Energy equality for the {N}avier-{S}tokes equations in weak-in-time {O}nsager spaces},
  journal = {Nonlinearity}, volume = {33}, pages = {1388--1403}, year = {2020}}

@Article{Tao16,
  author = {T. Tao},
  title = {Finite time blowup for an averaged three-dimensional {N}avier-{S}tokes equation},
  journal = {J. Amer. Math. Soc.}, volume = {29}, pages = {601--674}, year = {2016}}

@Article{LesShv18,
  author = {T. M. Leslie and R. Shvydkoy},
  title = {The energy measure for the {E}uler and {N}avier-{S}tokes equations},
  journal = {Arch. Ration. Mech. Anal.}, volume = {230}, pages = {459--492}, year = {2018}}

@Article{CafKohNir82,
  author = {L. Caffarelli and R. Kohn and L. Nirenberg},
  title = {Partial regularity of suitable weak solutions of the {N}avier-{S}tokes equations},
  journal = {Comm. Pure Appl. Math.}, volume = {35}, pages = {771--831}, year = {1982}}

@Misc{Uda26,
  author = {T. Uda},
  title = {Lean 4 formalization of {L}eray-{H}opf weak solution existence for the three-dimensional incompressible {N}avier-{S}tokes equations},
  howpublished = {\url{https://github.com/uda-lab/leray-hopf}}, year = {2026}}

@Incollection{CheFriShv10,
  author = {A. Cheskidov and S. Friedlander and R. Shvydkoy},
  title = {On the energy equality for weak solutions of the 3{D} {N}avier-{S}tokes equations},
  booktitle = {Advances in Mathematical Fluid Mechanics}, publisher = {Springer}, pages = {171--175}, year = {2010}}

@Article{Shv10,
  author = {R. Shvydkoy},
  title = {Lectures on the {O}nsager conjecture},
  journal = {Discrete Contin. Dyn. Syst. Ser. S}, volume = {3}, pages = {473--496}, year = {2010}}

@Article{NecRuzSve96,
  author = {J. Ne\v{c}as and M. R\r{u}\v{z}i\v{c}ka and V. \v{S}ver\'ak},
  title = {On {L}eray's self-similar solutions of the {N}avier-{S}tokes equations},
  journal = {Acta Math.}, volume = {176}, pages = {283--294}, year = {1996}}

@Article{Tsa98,
  author = {T.-P. Tsai},
  title = {On {L}eray's self-similar solutions of the {N}avier-{S}tokes equations satisfying local energy estimates},
  journal = {Arch. Ration. Mech. Anal.}, volume = {143}, pages = {29--51}, year = {1998}}

@Article{ChaWol20,
  author = {D. Chae and J. Wolf},
  title = {Energy concentrations and {T}ype {I} blow-up for the 3{D} {E}uler equations},
  journal = {Comm. Math. Phys.}, volume = {376}, year = {2020},
  note = {arXiv:1706.02020}}

@Article{Kuk06,
  author = {I. Kukavica},
  title = {Role of the pressure for validity of the energy equality for solutions of the {N}avier-{S}tokes equation},
  journal = {J. Dynam. Differential Equations}, volume = {18}, pages = {461--482}, year = {2006}}

@Misc{Ser24,
  author = {G. Seregin},
  title = {Remarks on {T}ype {II} blowups of solutions to the {N}avier-{S}tokes equations},
  howpublished = {arXiv:2304.04045; Commun. Pure Appl. Anal. (2024)}, year = {2024}}

@Misc{Hua26,
  author = {H. Huang},
  title = {Endpoint energy atoms force local pressure concentration in three-dimensional {N}avier-{S}tokes flow},
  howpublished = {arXiv:2608.30715}, year = {2026}}

@Misc{OpenAI26NS,
  author = {{OpenAI}},
  title = {Finite time blowup for {N}avier-{S}tokes},
  howpublished = {\url{https://cdn.openai.com/pdf/32d9f210-8b73-45e0-91bc-82a30aef8a9a/navier-stokes.pdf}; announcement \url{https://openai.com/index/navier-stokes-solution/}; Lean formalization \url{https://github.com/openai/NavierStokesAndEuler}},
  note = {Accessed 8--9 September 2026},
  year = {2026}}

@Misc{OpenAI26Euler,
  author = {{OpenAI}},
  title = {Finite time blowup for the {E}uler equation},
  howpublished = {\url{https://cdn.openai.com/pdf/315b36cd-ec98-4023-8342-93345194ece1/euler.pdf}},
  note = {Accessed 8 September 2026},
  year = {2026}}

@Misc{AlpBuc26,
  author = {L. Alp\"{o}ge and T. Buckmaster},
  title = {Blowup for the {E}uler equations with smooth forcing},
  howpublished = {preprint dated 7 September 2026, \url{https://cims.nyu.edu/~tristanb/euler.pdf}},
  note = {No author line; authors as named in the OpenAI announcement of 8 September 2026. Accessed 9 September 2026},
  year = {2026}}

@Incollection{Fef06,
  author = {C. L. Fefferman},
  title = {Existence and smoothness of the {N}avier-{S}tokes equation},
  booktitle = {The {M}illennium {P}rize {P}roblems},
  editor = {J. Carlson and A. Jaffe and A. Wiles},
  publisher = {Clay Math. Inst. and Amer. Math. Soc.},
  pages = {57--67},
  year = {2006},
  note = {Official problem statement, \url{https://www.claymath.org/wp-content/uploads/2022/06/navierstokes.pdf}}}

@Article{Ons49,
  author = {L. Onsager},
  title = {Statistical hydrodynamics},
  journal = {Nuovo Cimento (9)}, volume = {6}, number = {Suppl. 2}, pages = {279--287}, year = {1949}}

@Article{Eyi94,
  author = {G. L. Eyink},
  title = {Energy dissipation without viscosity in ideal hydrodynamics. {I}. {F}ourier analysis and local energy transfer},
  journal = {Physica D}, volume = {78}, pages = {222--240}, year = {1994}}

@Article{ConETit94,
  author = {P. Constantin and W. E and E. S. Titi},
  title = {Onsager's conjecture on the energy conservation for solutions of {E}uler's equation},
  journal = {Comm. Math. Phys.}, volume = {165}, pages = {207--209}, year = {1994}}

@Article{Ise18,
  author = {P. Isett},
  title = {A proof of {O}nsager's conjecture},
  journal = {Ann. of Math. (2)}, volume = {188}, pages = {871--963}, year = {2018}}

@Article{BucDeLSzeVic19,
  author = {T. Buckmaster and C. {De Lellis} and L. {Sz\'ekelyhidi, Jr.} and V. Vicol},
  title = {Onsager's conjecture for admissible weak solutions},
  journal = {Comm. Pure Appl. Math.}, volume = {72}, pages = {229--274}, year = {2019}}

@Article{DriEyi19,
  author = {T. D. Drivas and G. L. Eyink},
  title = {An {O}nsager singularity theorem for {L}eray solutions of incompressible {N}avier-{S}tokes},
  journal = {Nonlinearity}, volume = {32}, pages = {4465--4482}, year = {2019}}

@Article{BruDeL23,
  author = {E. Bru\`e and C. {De Lellis}},
  title = {Anomalous dissipation for the forced 3{D} {N}avier-{S}tokes equations},
  journal = {Comm. Math. Phys.}, volume = {400}, pages = {1507--1533}, year = {2023}}

@Article{BruColCriDeLSor24,
  author = {E. Bru\`e and M. Colombo and G. Crippa and C. {De Lellis} and M. Sorella},
  title = {Onsager critical solutions of the forced {N}avier-{S}tokes equations},
  journal = {Commun. Pure Appl. Anal.}, volume = {23}, number = {10}, pages = {1350--1366}, year = {2024}}
\end{document}